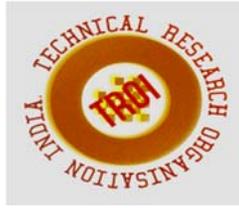

# MAPREDUCE SCHEDULER: A 360-DEGREE VIEW

Rajdeep Das[1], Rohit Pratap Singh[2], Ripon Patgiri[3]
Department of Computer Science & Engineering
National Institute of Technology Silchar, Assam, India
Email:radsria@gmail.com[1], rohitkako@gmail.com[2], ripon@cse.nits.ac.in[3]

**Abstract**
**Undoubtedly, the MapReduce is the most powerful programming paradigm in distributed computing. The enhancement of the MapReduce is essential and it can lead the computing faster. Therefore, there are many scheduling algorithms to discuss based on their characteristics. Moreover, there are many shortcoming to discover in this field. In this article, we present the state-of-the-art scheduling algorithm to enhance the understanding of the algorithms. The algorithms are presented systematically such that there can be many future possibilities in scheduling algorithm through this article.**

**In this paper, we provide in-depth insight on the MapReduce scheduling algorithm. In addition, we discuss various issues of MapReduce scheduler developed for large-scale computing as well as heterogenous environment.**

**Index Terms: Scheduler, MapReduce, MapReduce Scheduler, Large-scale computing, Data-intensive computing.**

## I. INTRODUCTION

Unquestionably, the MapReduce is the most powerful and popular programming paradigm of parallel and distributed processing engine. The MapReduce meets very vast range of application for large-scale processing. The MapReduce is proven to be the best programming platform for dataintensive computing. The Google File System (GFS) was introduced in 2003 [31] and the MapReduce was introduced by IT giant Google Inc. in 2004 [32]. It was absolutely proprietary [31] [32] and written in C++. The Yahoo! company wanted the replica of the MapReduce and Hadoop [2], has got huge popularity. The MapReduce is very easy programming framework to develop the largescale distributed application within a few lines of code. The MapReduce shows the properties of simple yet powerful. However, the gigantic scale of data processing is not a piece of cake and it has become easier on the MapReduce programming environment. However, the MapReduce has certain limitation which is highlighted later section.

The MapReduce is developed under the Apache Hadoop framework [2] and it is proven to be the best framework for application development in large-scale distributed computing environment. However, the large scale computing is split into several small pieces of computing and distributed the task in several computing nodes. The tasks are scheduled in different suitable node.

Nevertheless, the detecting or finding the suitable node for running a task is a very difficult job. Further, the suitable node must enhance the job performance and hence, this is the most relevant algorithm to fine tune. There is always a scope to enhance job performance by placing the tasks in suitable node. The most common scheduler is first-come first-serve (FCFS). However, the Hadoop MapReduce has priority, capacity and fair scheduler. Interestingly, these three schedulers are not enough to enhance the performance of jobs. There are many parameters to be considered, for instance, data locality.

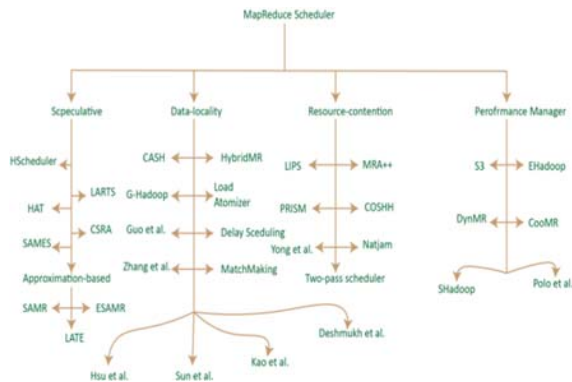

Fig 1: Taxonomy of MapReduce scheduler

## II. SPECULATIVE SCHEDULING ALGORITHMS

The most of task submits on time excepts one or two. Albeit, the fast processing of all other tasks, one task can lengthen the job completion time and it is known as Achilles' Heel of MapReduce. There are many reasons for a task to be a straggler, namely, faulty hardware, wrong configuration, network congestion, link failure, low processing power, and heavy background noise. The straggler task should be identified as early as possible so that speculative execution of the task can be started. A performance can also be degraded when aggressive speculation of a task in testing whether the task belongs to straggler or not.

### A. *Approxiation-based Scheduler*

The Longest Approximate Time to End (LATE) is the milestone for MapReduce scheduling algorithm which was introduced in 2008[1]. The focal point of this algorithm is speculative execution of the straggler task to increase MapReduce performance. LATE[1] algorithm performs speculative execution of a task in the heterogeneous environment. The key point of LATE is prioritizing the task to speculate on the different node. Interestingly, LATE selects fastest node to launch the task speculatively and caps the speculated task to prevent thrashing. Whenever a node is requesting for a task and has its slot free, Hadoop schedules the failed task with high priority. If no such task present, then LATE pick a non-running task from a job to schedule in Hadoop. Hadoop uses a progress score for each task which is 0 to 1 to select the task to be run speculatively. In map phase, the $prgress\_score$ is the fraction of input data read to the total data whereas the $prgress\_score$ for reduce phase is split into three phases, namely, Copy phase, Sort phase and Reduce phase. The key idea of LATE scheduler is that it always speculatively execute the task which will finish late in near future. So, it improves the performance and shorten the response times. LATE uses progress of the map and reduce task to approximate the completion time of the task in the future.

LATE estimates the progress rate of each task as $ProgressRate = \frac{progress\_score}{T}$, where $T$ is the task running time. And completion time for every task can be calculated using the formula of $TimeToComplete = \frac{(1-progress\_score)}{ProgressRate}$.

However, Hadoop launches task based on data locality whereas LATE doesn't consider locality while launching the speculative map task. Assumption used by LATE is that most of the map task is data local, therefore, network utilization is less. When it launches non local task, data can be sent to the node using the network. Moreover, It launches the backup task for inappropriate tasks. LATE considers the slow nodes using the progress rate, but it doesn't consider whether a map task or a reduce task is slow. In heterogeneous environment the value of a node changes with the execution of map and reduce task, with changes in input the $prgress\_score$ for the map and reduce task differs in different nodes.

In a heterogeneous environment, LATE suffers because of the static nature. However, a Self-Adaptive MapReduce scheduling (SAMR) algorithm[5], which computes $prgress\_score$ of tasks dynamically and reconciles to the constantly diverse environment. It works well compared to LATE in a same manner, like it uses historical information stored in the node for determining the slow task which needs to be executed speculatively. Dynamic nature of SAMR significantly improves the MapReduce execution time and system resources as shown in paper[5]. SAMR saves system resources by classifying slow task further in slow map task and slow reduce task. SAMR works on the key insight that slow task prolongs the execution time of the whole job and takes more time to finish a job because of the differences in computational power. SAMR[5] calculates more accurate progress score compare to LATE because it considers historical information stores in the node. The historical information is saved on each node in xml format to update the historical information of a node. To start with, historical information of the node is get by the TaskTracker. Then, based on the historical information the parameter for the task gets tuned. SAMR finds the task from the queue of slow task and schedule it over the fast TaskTrackers which are freely available to achieve high performance and response time.

However, SAMR has some drawbacks. It suffers with different map and reduce stage weights of different types of jobs. As well as for the same type of jobs when the input data set changes the stage weight changes. SAMR doesn't consider the change in weight of the stage dynamically. In 2012, an Enhanced Self-Adaptive Map Reduce [ESAMR] [25] scheduling algorithm is proposed to enhance the speculative re-execution of straggler tasks in MapReduce. Using a k-means clustering algorithm, it offers historical stage weight information on each node to identify slow task node with high veracity, and divide them into k clusters. It arranges the task into several clusters and job execution time on the node is calculated based on the cluster's weights.

### III. LOCALITY-AWARE REDUCE TASK SCHEDULER (LARTS)

LARTS[9] is the only scheduling algorithm for Reduce task based on data locality. Native Hadoop scheduler works on the phenomenon of early shuffle of reducers near the partitions, it helps in improving the performance, but it simultaneously increases the network traffic. Regardless, LARTS maintains the aid of interleaving process with less network traffic. Early shuffle mechanism in Hadoop improves performance by scheduling a reduce task before intermediate data present in each partition is available. Generally, after the commitment of certain percentage(5%) of mappers, scheduling process of reduce task starts. The reason for early shuffle is to provide mappers intermediate output partition to a corresponding reducer early so that it doesn't have to wait for getting processed by reducers, and it helps to improve turnaround time for MapReduce job. LARTS algorithm modifies Hodoop to distinguish the network traffic into node local, rack local and off rack. When a reducer is scheduled to the node having an intermediate data partition to be consumed by the same reducer is known as node local traffic, i.e., node local traffic is negligible because reducer is scheduled to the same node containing the map partition. Rack local traffic is obtained when the partition generated by map task of the node is shuffled to reducer which is scheduled to some other node present in the same rack. Whereas in the off rack traffic the partition is shuffled to the reducers which is scheduled in off rack node. Hadoop uses the resource monitor to estimate the input size for reducers before it gets scheduled in the TaskTracker. If the Hadoop job tracker finds the TaskTracker which is requesting for reducers, doesn't have enough space for reducer then job tacker will not schedule the reducer to that particular TaskTracker (TaskTracker is rejected). Early shuffle is the feature provided by the native Hadoop scheduler, reducers get scheduled with the mappers running on the node. Mappers which are busy generating the largest amount of intermediate output need more resources, in such scenario, TaskTrackers holding busy mappers make a request for reducers is rejected by job tracker. Subsequently, the intermediate output of the busy mappers has to be scheduled to the reducers present in some other node and it increases the traffic in the network. Therefore, job tracker schedules the reducers to the TaskTracker holding busy scheduler in order to reduce the network traffic of shuffling the intermediate data of large partition.

LARTS is used to schedule reduce task to the TaskTracker so as to reduce the network traffic for large shuffled data. MapReduce framework is well aware of the memory location of the input data for scheduling the map task. Whereas, location of the partition generated by the map task is not known to the framework, i.e., scheduling reduce task is a challenge. Similar to the map task, algorithm makes the MapReduce aware of network location for scheduling, reduce task based on data locality. Network location of the intermediate dote will be fully known when the all map tasks in the TaskTracker complete. Early shuffle of reduce tasks helps in increasing efficiency and decreasing the job turnaround times. Early shuffle can be initiated after a defined number of map task finishes their execution. A sweet spot is defined in the program where early shuffle can be initiated. In addition to the network location of each partition, a MapReduce framework should keep track of the size of all partitions in order to schedule reduce task based on data locality.

Reducer has to be scheduled to the mapper generates the partition and better data locality can be accomplished by scheduling the reducer at the TaskTracker hosting the partition. Maximum-rack of reducer is defined by the number of the rack that holds the maximum number of partitions whose accumulated size is larger than any other accumulated size of other racks holding the partition for the same reducers. In the same way, the algorithm defines maximum-node of reducer as the node contains the biggest partition for reducer at the maximum-rack. Maximum-rack and maximum-node is calculated using the network location and size of partition. In this approach, some problem arises, namely, scheduling delay, scheduling skew, lower parallelism and poor throughput. A scheduling delay occurs when the TaskTrackers request for reducers simultaneously and gets rejected because they fail in the criteria of maximum-rack and maximum-node and this situation leads to scheduling delays. Whereas, scheduling skew occurs because of the variance in the TaskTracker being preferred by the reducers. Very poor parallelism and poor throughput results after a skew in scheduling. LARTS deals with the scheduling delay, scheduling skew, system utilization and parallelism by fragmenting the reduce task among the TaskTrackers requesting for the job. Mechanism to deal with the above mentioned shortcomings is rejection counter, Rejection counter is present for every TaskTracker, it increments whenever a TaskTracker gets rejected by the job tracker. Threshold is present for restricting the number of times a TaskTracker gets rejected after that job tracker assign reduce task to the requesting TaskTracker.

## IV. DATA LOCALITY SCHEDULING
### A. A Data Distribution-Aware Scheduling Algorithm

A data distribution aware task scheduling algorithm [4] is the foremost algorithm proposed based on data locality. The key point is scheduling a task where data is available locally. So network communication and data transmission rate can be reduced if job tracker schedule the task to the TaskTracker having the data. Since, moving computation is cheaper than moving data. In data distribution aware scheduling task is distributed on the slave nodes based on data locality. Scheduling is done based on the scheduling priority of each task and the node requesting the task. Calculation of scheduling priority is the Main module of Master Node.

The priority scheduler for a task is calculated based on three factors, namely, *Replica-num*, *WorkerTasklist*, and *Task-waittime*. First, the *Replica-Num* determines the number of map tasks those can be executed the local task, i.e., when a task will be assigned to TaskTracker, the data for that task should be present locally to that node. Data is replicated by HDFS [2] over the different nodes present in the network. So while considering the priority, it considers less priority for the task which is having high *Replica-Num*. After that, *WorkerTasklist* determines the length of the localize task list for the nodes saving the data to be executed by the map task. If the length of localize task is more, then the scheduling priority of that task is high. At last, *Task-waittime* determines the time period of a task from the starting to the current moment of time. So, task whose waiting time is high has the highest scheduling priority.

Now, computing the scheduling priority for each node requesting for the task. Slave node request for task through a heartbeat signal and advertise free slots. Scheduling priorities of workers are calculated using three factors Worker's Local Task list, *Request-Num* and *HistoryInfo*. So first of all, Workers Local Task list is maintained for each node. Task present in the list can be executed by each worker locally. If node local task list is not empty, then scheduling priority can be set at maximum. And master schedules task in this list of the worker first. Second, *Request-Num* determines the number of requests since the node gets the last task. It means a node request for task whenever a node is free in every heartbeat signal. Count of *Request-Num* shows that the worker has less load and can manage more task. So, bigger the Request-name has higher priority. At last, h*istory-Info* gives the success rate for task processing on the node. Scheduling priority of worker is proportional to the *History-Info*. With this intention, task getting scheduled over the node corresponds to the workers, local task list, if the worker local task list has some task then it has the highest priority. Else the other two factors will come into the picture.

### B. Delay Scheduling Algorithm

Delay scheduling algorithm is a very simple technique for accomplish locality and fairness in MapReduce scheduling. Fairness can be achieved by fair sharing of resources, but strict establishment of fair sharing adjust locality, because when a request generated by a node might not have the data for the task to be scheduled next. Hence delay scheduling [6] relaxes fairness slightly, in which job waits for a limited amount of time before getting scheduled on a node that has data for it. Problems related data locality while considering fairness is Head of line scheduling and Sticky slots.

Head of line scheduling is the data locality issue arises because of small jobs having tiny input files leads to less number of input blocks, so in fair sharing scheduling, tiny job having a minimum number of tasks, next task is scheduled on the free slot requesting for the task, and it does not consider the node to launch the task. However, in sticky slots problem occurs even for large job if fair sharing is used. The problem is that a slot can be repeatedly assign for the same job. The problem arises when jobs are sorted according to the increasing order of running tasks, one task gets scheduled on one of the slot and it finishes. Then the next task of the same job will be scheduled back to same slot. In consequence, it leads to the situation that jobs never leave their original slots. The main idea behind delay scheduling is that, even though a slot is free and requesting for task irrespective of the fact that it contains the data for that task, it will be unlikely to assign task to that slot because tasks finishes so quickly that some slot with the data for it will free up in the next few seconds. So delay scheduling along with fair sharing solves the problem of data locality.

Fair sharing with delay scheduling states that set the job with a value $D$, scheduling a non local task in the job will be skipped $D$ times until it gets the local task, and once it crosses the limit of $D$ times then it let the job launch arbitrarily many non-local tasks without resetting the count of $D$. However, once it manages to launch a local task, then the value will be set back to 0. Two levels of delay scheduling are also proposed, it states that first it delay the job by $W_1$ seconds to find a node local task, and then it waits till $W_2$ seconds to find a rack local task. It achieves more level of data locality rate compare to the single level of delay scheduling because here it is considering the fact that data can be accessed through rack locality. Initially, it checks for node local task for jobs, when it fails to find then it checks for rack local task. When it fails to find the node local task as well as rack local task, it launches arbitrarily non local task. Considering the value of $D$, it cannot be set higher because then it will lead to resource wastage for the non-local node requesting for the task. It cannot be set low because then it will work like general fair scheduling algorithm. It should be set ideally, means not too high and not too low.

Limitations of Delay scheduling is when a large portion of the tasks is much larger than the average job, then effectiveness will become less. It delays the task by $D$ times, but since larger job are assigned to the slots, within that $D$ times slots might not become free and it will lead to lower utilization.

### C. Data-locality Aware Task Scheduling Algorithm

Data locality is the concern of this [7] scheduling algorithm. It takes decisions based on the waiting time and transmission time of the task. In brief, the idea of this scheduling algorithm is that whenever a requesting node request for a task, the method selects the task from task list whose data is stored on the desired node. If no such task encountered, then the method picks a task from the list where data are near to the task. The basic idea behind calculating the waiting time and transmission time is that if the waiting time of selected task is less than the transmission time of the selected task, then the task will be reserved for the node having the data stored in it, otherwise it will schedule the task over the requesting node. The waiting time of a task stands for the remaining time needed to complete the executing task on the node having the data stored. Multiple task executes in a node simultaneously, hence the waiting time for a task having data stored on the node is represented by the time taken to finish the shortest task among all, once the shortest task will complete execution the requested task can be scheduled. The transmission time for the task stands for the time taken to copy the input data to the node requesting for task, time rely on network communication. Selection of task after getting the request from the requesting node is done based on the probability calculation when a node local task is not present. It maintains the different level of the task, when first level tasks do not fulfill the condition, it checks in second level and then followed by the third level. The probability gets calculated so that they can be scheduled according to their probability. Task with high probability gets chance to be scheduled first on the requesting node, whereas the task with low probability get reserved for the node storing their data.

### D. MatchMaking

The key idea of the matchmaking technique [8] is to assign tasks to a TaskTracker, the local map task is always preferred over other map task. To avoid network transmission rate the local task a preferred over non local task. Another technique used by matchmaking is a locality marker. MatchMaking marks the node and ensure that each node gets the fair chance to grab it local task. Scheduler grants the request based on the heartbeat signal generated by the requesting slave nodes, so data locality is randomly decided by the heartbeat sequence of slave node. In such scenario where large cluster is present with small jobs then data locality rate could be quite low.

The matchmaking scheduling main idea is to assign tasks based on locality. It makes sure that each slave node should get a fair chance to grab the task for which data is stored in the node before any task gets assigned to it. When a local map task is not found in the first job, then the scheduler will continue searching in the succeeding jobs. Whenever it finds a local map task, it schedule the map task over the node requesting for the task.

Slave nodes should get fair chance to grab the node local task, when a node fails to find a node local task, it doesn't launch the non-local task, else it waits for some time and request again for node local task. The main idea behind is, tasks finish very fast so when a node wait for one heartbeat signal, the other nodes also finish their execution and node local task gets available. To avoid wasting of resource by keeping the node ideal, the non-local task will be assigned to in the next heartbeat. Matchmaking scheduling allows a slave node to take at most one nonlocal task in every heartbeat. Matchmaking scheduling achieves high data locality and high cluster utilization. Matchmaking scheduling uses a Locality marker assigned to every slave node, it keeps track whether a local task is launched in the slave node within that heartbeat signal, if not then in next heartbeat signal a non-local task will be assigned to the node.

### E. Context-Aware Scheduler in Hadoop (CASH)

In MapReduce framework, Job scheduler is the key component as it decides and controls when and where a job task get executed. Context aware scheduler in Hadoop [10] design principles is that, firstly a large percentage of MapReduce jobs are periodic in nature. It means the largest number of jobs, having roughly the same characteristics and execute at the same time. Secondly, Hodoop cluster contains heterogeneous nodes that means the computation and disk capabilities of the nodes present in the cluster are not same. It happens because more nodes are added in cluster with time and old nodes are replaced, so the Hadoop cluster becomes heterogeneous in nature. The context of the node should be known before we schedule the job. Context aware scheduling knows the context of the job, job characteristics (CPU or I/O bound) and the resource characteristics computation or I/O strength of the node in the cluster.

CASH implementation in Hadoop can be done in few steps. Initially, Classify the jobs as CPU bound or I/O bound, CPU bound jobs are more computation and processing oriented where as I/O bound jobs are Input output oriented. Job classification is done using the summary logs, whenever a new job runs for the first time and finish execution a log is maintained. From logs it can be obtained that whether Map Reduce I/O rates are less than the Disk I/O rate, then it can be classified as CPU bound job, otherwise, it can be classified as an I/O bound job. Secondly, Classify the nodes as computational or I/O good. Classification of node can be done by running the CPU and I/O benchmark on each node respectively. If the CPU benchmark is more compared to the I/O benchmark for a node than that node is classified as computational good or else classified as I/O good. Lastly, scheduling the task of the job with different demands to the node which can meet the demands, CPU bound job to the computational oriented node where as I/O bound job to the I/O oriented node.

The CASH scheduler schedules task of jobs based on the both data-locality and requirement match. While classifying the node, some node appears to be average in CPU and Disk then it schedule jobs randomly based on FIFO order. Initially job is executed in FIFO order to get the job summary logs, to classify the jobs as CPU bound or I/O bound. In the preferred algorithm, the scheduler searches for the Map task which fulfills both locality and the requirement of the node. If both criteria fulfill then schedule the map task to that node. Else if no data local task found, then scheduler search for the non-local task with the match requirement. Whereas for reduce task scheduling it does not consider the data locality, and searches for the task fulfilling the node requirements. Sometimes requirement does not match, then it schedules task based on data locality. If nothing matches, then it schedule task based on FIFO order.

## V. RESOURCE-CONTENTION SCHEDULING

A. *Towards a Resource Aware Scheduling*

In the Hadoop cluster (2009) [3], many users submit the job simultaneously because the Hadoop cluster is distributed in nature. This algorithm deals with the improving resource utilization when different kinds of user submit jobs on the cluster. Generally, Hadoop scheduler is not aware of the nature of the job submitted and Hadoop scheduler prefers to run a map task of the job present on the top of the job queue. Job tracker assigns tasks to the TaskTrackers when TaskTracker has a free slot and request for a task. So, in simple Hadoop job tracker assigning tasks to the TaskTracker has some limitations. Firstly, the allocation of the task to the node having data is done by the job tracker without considering the workload of the node and the availability of the node. Secondly, when the task is assigned to TaskTracker, and job running on TaskTracker becomes slow, then job tracker cannot keep track of such task.

TaskTracker tracks the resources in Hadoop has some weakness. Firstly, it fails to monitor the capacity and load level of individual resources present in TaskTracker nodes. Secondly, it fails to utilize the available resource matrices guide to take decision for scheduling task. An improved scheduler [3] is proposed to overcome the above mentioned drawbacks. It monitors the resource available in every node and takes the decision based on the resource matrices. TaskTracker resource monitoring is used to monitor the resources in the TaskTracker level, Each TaskTracker monitors resource in the node such as CPU utilization, disk channel, I/O and number of page faults per unit time for the memory system.

In job tracker it uses resource matrices to schedule task into the node requesting for tasks. Firstly, Dynamic free slot advertisement, it means the TaskTracker node contains a fixed number of available task slots, but using resource matrices it computes the free slots dynamically. Secondly, free slot priorities, in this mechanism the fixed number of computational slots present in every node is retained and instead the order in which TaskTracker node will advertise its free slots will be decided based on their resource availability. Lastly, Energy efficient scheduling, the energy consumption in the scheduler is less than the consumption of energy in the Hadoop basic scheduler.

B. *Natjam*

Natjam system [15], provides random job priorities, hard real-time scheduling and efficient preemption for MapReduce those are resource constrained. In the MapReduce framework jobs comes with difficult priority, high priority job (short completion time) and low priority jobs (long completion time) the traditional way to solve the problem of priority is setting different clusters, which later lead to the inefficient resource utilization and long job finish time.

The algorithm aims, 1. To run all the types of job, regardless of priority, completion time in the same MapReduce cluster. 2. Attain a lowest completion time for higher priority jobs. 3. Optimizing completion times of lower priority jobs. The mentioned goals are attained using Natjam algorithm.

C. *CooMR*

The available slots in the processor are represented as free map and reduce slots, and assign them to different tasks. The cross-task coordination CooMR is designed for efficient data handling in MapReduce programs. The CooMR [19] leads to increase task coordination, enhance system resources throughput and significantly enhance the process time of MapReduce jobs. MapReduce programs face two key performance issues, namely, task interference and excessive I/O. Task interference can cause prolonged execution time for map tasks. Whereas excessive I/O can degrade the disk I/O bandwidth.

D. *A Load-aware Scheduler*

Dynamic loading [26] of job always been an issue in the Hadoop scheduler, so a new scheduler load aware scheduler is proposed. It comprises of two modules, namely, data collection module and the task assignment module. Data collection module gathers the system level information periodically from the TaskTracker. Whereas task assignment module makes the scheduling decision according to the TaskTrackers info collected by the data collection module.

E. *MOMTH*

The multiobjective considered in this algorithm [20] is related to the user, resources and with restraints like deadline and budget. Transformation of sequential task is needed to lessen the execution time & to diminish the resource requirement. The main difference between the single objective and multiobjective approach is the requirement condition and the alteration of an objective in the fields. The MOMTH algorithm has two main objectives. First, having an optimal work assignment in the cluster and avoiding resource contention. Second, to complete tasks within budget and meet deadlines.

F. *CSRA*

Data skewness is the one of the major reason of straggler emergence and makes data assignment to reducer imbalance. The main objective of CSRA [27] is to reduce execution time and coefficient of variation by changing the order of task list and dividing the big clusters. CSRA has less overheads and increase the execution time of applications. CSRA (cluster splitting based on resource allocation algorithm) mainly focuses on fixing the problem with data skew.

E. *Dynamic Reduce Task Adjustment for Hadoop Workloads*

The challenging aspect [28] of executing the Hadoop job is the management of reduce task. The algorithm proposes, first an approach for calculating the appropriate number of reduce tasks per job. Second, usage for profile job in gathering information for the reduce task computation and third, two different policies for fragmenting the reduce task to the available system resources when multiple jobs execute concurrently in the cluster. Appropriate number of reduce task is calculated using the historical information regarding the input data sets.

## VI. PERFORMANCE MANAGING SCHEDULER

A. *Joint Optimization of Overlapping Phase in MapReduce*

This algorithm [14] deals with the problems occur when map & reduce phase, execute in an overlapping manner. MapReduce phase like Map, shuffle \& reduce phase works in such a way that the output of one phase becomes the input to another phase. So, it becomes challenging to schedule and allocate resources in complex and critical MapReduce system. It uses shortest remaining processing time (SPRT) first scheduling algorithm minimizes average response time within a single server. So, map task with large shuffle size completed first to avoid wastage of resources.

B. *Busrtiness-aware I/O Scheduler*

Virtual environments like cloud computing and virtual cluster are popular recently because of the low cost and flexibility. In this algorithm [16] a burstiness-aware I/O scheduler is proposed. Long seek distances in a disk leads to I/O interference and then it leads to numerous context switches in the virtualization software. For utilizing the I/O bandwidth without interference, the proposed scheduler encounters I/O burstiness of a virtual machine on-line. The key idea of the scheduler is to schedule the identified bursty virtual machines with a comparatively big time quantum in round robin fashion and to avoid the starvation situation I/O bandwith for the non-bursty virtual machine is used.

### C. DynMR

DynMR algorithm [17] is used to improve the performance of MapReduce, it lists some problem present in the existing MapReduce implementation. 1. Optimal performance parameter is challenging to select for a single job in a dedicated environment. In multijob cluster it lacks the efficiency to construct parameters that can behave optimally. 2. Long job execution leads to taileffect. 3. Hardware resources in inefficiently used. DynMR uses the interleave way of execution where several partially completed reduce tasks and map tasks executes. It consists of three components. 1. Detection of underutilized resources in the shuffle phase and give up the allocated hardware resources efforlessly to the next task. 2. All the reduce task is assembled in progressive queue and execute in interleave rotation. 3. Merges the threads of all partial complete reduce task, it allows to keep the data segment of multiple reduce task in one JVM heap.

### D. Throughput Driven Scheduler

The native Hadoop scheduler are not job-intensive scheduler so provided less throughput. So the algorithm proposes a way for improving throughput using jobintensive scheduling. A novel job scheduling technique is proposed, Throughput driven [29] task scheduler for obtaining high system throughput in the job-intensive MapReduce environment. The algorithm summarizes several aspects which can impact throughput of a job intensive MapReduce environment. High ratio is attained for local task assignment using throughput driven task scheduler, and full advantage of the system resource can be attained effectively.

### E. ARIA

The issue in shared MapReduce cluster [30] is to keep track of the resource allocation of different applications for achieving their performance goal. With native Hadoop scheduler, no such job scheduler is present that can appropriately allocate the resources to the job once given a job completion deadline. The proposed algorithm called AIRA deals with the above mentioned problem. ARIA consists three interrelated components. Initially, a job that is frequently executed on a new dataset, job profile is maintained which contains the info about Map and Reduce tasks. Lastly, the MapReduce performance model is made for estimating the amount of resources for job completion.

## VII. RECENT SCHEDULING ALGORITHMS

MapReduce scheduler is the emerging area for many researchers, scheduling algorithm based on data locality, performance manager, speculative task execution and resource contention is as follows.

### A. Job aware scheduling algorithm for MapReduce framework.

The approach [11] is proposed to decrease the runtime of the MapReduce jobs running in the cluster. When some task running on the node and other task appears, then scheduler tries to allocate the task to the node if it doesn't affect running one. The algorithm selects the compatible task from the pending list of the task to the already running task in the Node. The proposed algorithm achieve low runtime by avoiding the overload to a node. The proposed algorithm tries to monitor the usage of resources in each task and each node. Using the intelligent scheduling the algorithm tries to maintain stability at node and cluster level.

### B. Resource-aware adaptive scheduling for MapReduce cluster

In the Hadoop native scheduler, the static number of execution slots is examined to determine the capacity of the cluster. But the native scheduler fails to get the individual requirement of the each job. The resource aware adaptive scheduler [12] gets the job requirement through the job profiling and adjust the number of slots on each machine. The capacity calculation in the Hadoop cluster is the function of the task can run concurrently in the system.

### C. Improving MapReduce performance in heterogeneous network environments and resource utilization

In the MapReduce framework when some slots are idle and task is scheduled based on data locality, the resource reserved for the task will be wasted if no such task present at that moment. The algorithm propose [13] the concept of resource stealing, which enables running task to steal utilized resource and return when new task will be assigned. It helps in increasing overall job execution time and resource utilization.

D. *A case study of MapReduce speculation for failure recovery*

The existing Hadoop framework has some major drawback which hinders the efficiency of job execution during the recovery of failure, which is termed as speculation breakdown. So, failure aware speculation scheme is suggested to solve the problems related to speculation.

To overcome the problem of speculation mechanism, the proper examination of the existing speculation mechanism is done. First [18], it examines the impact and implication of failure in node using the current speculation mechanism. Second, it introduces new speculation scheme FARMS. Last, it uses a fast analytic scheduling algorithm to work with YARN, which adds resilience to the heterogeneous real-world environment.

E. *Map task scheduling in MapReduce with data locality: Throughput and Heavy-Traffic optimality*

MapReduce program shows high performance and utilization when the task present in the program scheduled based on data locality. Some limitations present in the MapReduce cluster while scheduling based on data locality. Traffic optimal algorithm [21] (2016) is proposed along with that it focuses on the right balance between data locality and load balancing to simultaneously maximize throughput and minimize delay.

Based on the stochastic process model the map task arrives at the beginning of each time slot will be scheduled. The algorithm also provides support for backlogged tasks.

F. *HScheduler*

On MapReduce cluster, the execution time of the job depends on the way the map task is scheduled, the overall make span and the resource utilization depends on the scheduling of the map and reduce tasks. The goal of HScheduler [22] is to design the MapReduce scheduler, which reduces the make span of the jobs.

So, the algorithm provide a new modeling and scheduling approach for multiple MapReduce jobs, proposed online and offline both type of scheduling approach.

G. *Shortest remaining time first policy (STRF)*

Native Hadoop [23] scheduler faces the problem of how to reduce makespans by reducing job waiting time and execution times. Native scheduler, improves execution time by considering waiting time. STRF algorithm is proposed in the shared Hadoop cluster shortest-remaining time first (SRTF) in shared Hadoop cluster, estimates the remaining time of a job and preempt job whenever needed.

F. *Towards efficient resource provisioning in MapReduce* The key objective of the algorithm [24], for any work assigned in Hadoop MapReduce is to attain optimal number of task resources. Algorithm proposes the standard method for calculating the optimal number of tasks for any work assigned in Hadoop MapReduce. For optimal resource provisioning for any MapReduce workload, the algorithm develops a job profiling method of getting the runtime samples of the cluster. The algorithm provides stepwise computation processes with a mathematical formula for the runtime graph function. The algorithm is designed for best trade-off point for a work assigned as input and output the exact recommended number of task resources.

## VIII. CONCLUSION

Finally, we conclude that many works have been done so far to enhance MapReduce execution time. Moreover, we have discussed a small change can lead to drastically change in MapReduce performance. We have also discussed many issues related MapReduce scheduler. Furthermore, this article provide insight on the recent development of MapReduce scheduler.

## REFERENCES

[1] M. Zaharia, A. Konwinski, A. D. Joseph, R. Katz, and I. Stoica, ``Improving MapReduce performance in heterogeneous